\def\rmd{\mathrm d}

\def\etal{et al. }

\documentclass[reprint,aip,numerical,amsmath,groupedaddress,fleqn]{revtex4-1}

\usepackage[mathlines]{lineno}
\modulolinenumbers[5]
\usepackage{graphicx,bm}

\begin{document}

\title[Relationship between electron density and effective densities]{Relationship between electron density and effective densities of body tissues for stopping, scattering, and nuclear interactions of proton and ion beams}

\author{Nobuyuki~Kanematsu}\email{nkanemat@nirs.go.jp}
%\affiliation{Department of Quantum Science and Energy Engineering, Tohoku University, Aramaki-Aza-Aoba 01, Aoba-ku, Sendai 980-8579, Japan}

\author{Taku Inaniwa}
\author{Yusuke Koba}

\affiliation{Department of Accelerator and Medical Physics, Research Center for Charged Particle Therapy, National Institute of Radiological Sciences, 4-9-1 Anagawa, Inage-ku, Chiba 263-8555, Japan}

%\date{\today}

\begin{abstract}
\begin{description}
\item[Purpose]
In treatment planning of charged-particle radiotherapy, patient heterogeneity is conventionally modeled as variable-density water converted from CT images to best reproduce the stopping power, which may lead to inaccuracies in the handling of multiple scattering and nuclear interactions.
Although similar conversions can be defined for these individual interactions, they would be valid only for specific CT systems and would require additional tasks for clinical application.
This study aims to improve the practicality of the interaction-specific heterogeneity correction.
\item[Methods]
We calculated the electron densities and effective densities for stopping power, multiple scattering, and nuclear interactions of protons and ions, using the standard elemental-composition data for body tissues to construct the invariant conversion functions.
We also simulated a proton beam in a lung-like geometry and a carbon-ion beam in a prostate-like geometry to demonstrate the procedure and the effects of the interaction-specific heterogeneity correction.
\item[Results]
Strong correlations were observed between the electron density and the respective effective densities, with which we formulated polyline conversion functions. 
Their effects amounted to 10\% differences in multiple-scattering angle and nuclear-interaction mean free path for bones compared to those in the conventional heterogeneity correction.
Although their realistic effect on patient dose distributions would be generally small, it could be at the level of a few percent when a beam traverses a large bone.
\item[Conclusions]
The present conversion functions are invariant and may be incorporated in treatment planning systems with a common function relating CT number to electron density.
This will enable improved beam dose calculation while minimizing initial setup and quality management of the user's specific system.
\end{description}
\end{abstract}



\eprint{Revision to \href{http://link.aip.org/link/doi/10.1118/1.3679339}{Med. Phys. 39, 1016 (2012)}.}

\maketitle

\section{Introduction}

In the present practice of radiotherapy treatment planning, patient-specific material information is obtained from x-ray CT images.
The CT number given to each image pixel represents the x-ray attenuation coefficient of the material of that pixel, from which the effective density for the treatment beam is estimated on the assumption of one-to-one correspondence.
Taking water for a reference material, the effective density of a material is defined as the thickness ratio of water to the material for an equivalent dosimetric effect, which approximates to electron density for megavoltage photons or stopping-power ratio for charged particles.

The empirical relationship between CT number and effective density for a radiation of interest generally carries with it inaccuracies due to the limited tissue equivalency of sample materials.
Schneider \etal developed the stoichiometric method to construct improved relations for body tissues.\cite{Schneider \etal 1996}
They modeled the physics of x-ray attenuation using measured CT numbers of well-defined materials, with which they calculated the x-ray attenuations in the standard body tissues of known elemental composition, mass density, and electron density,\cite{ICRU 1989} in addition to their stopping-power ratios for protons, to construct the relations.

With such conversion, patients are modeled as variable-density water to assure range accuracy for charged-particle beams in conventional heterogeneity-correction algorithms.
As the body tissues deviate from water in elemental compositions, these algorithms only approximately address the other interactions.
Matsufuji \etal took the stoichiometric approach also for multiple scattering and nuclear interactions to study the validity of the conventional heterogeneity correction.\cite{Matsufuji \etal 1998}
They found that there would be 10\%-level errors in multiple-scattering angle and nuclear-interaction mean free path of protons in bones although their effects on patient dose distributions would be small in realistic situations.
Nevertheless, these definite errors may be corrected by the interaction-specific density conversions.
In fact, Szymanowski and Oelfke proposed a specific CT-number conversion for multiple scattering.\cite{Szymanowski and Oelfke 2003}
Palmans and Verhaegen proposed another CT-number conversion for proton nuclear interactions.\cite{Palmans and Verhaegen 2005}

Those studies offer valid methods for improved heterogeneity correction, which however have to be applied to individual scanning conditions of individual CT systems.
The resultant conversion tables must be separately set up and selectively used in treatment planning systems with periodical constancy tests or updates, which would greatly complicate the system-management tasks in the clinical environment.
The purpose of this study is to establish a simpler procedure and to provide necessary data for equivalently accurate conversion.
We take electron density for a reference quantity and similarly construct conversions for the effective densities of body tissues as a function of electron density.
As the conversion from CT number to electron density is commonly available in treatment planning systems, implementation of the invariant conversion functions in algorithms will offer accuracy improvement without further complicating data management.

\section{Materials and Methods}

\subsection{Electron density}

We used 92 of the ICRU body tissues,\cite{ICRU 1992} excluding obsolete, extreme, or artificially extracted materials such as ICRU-33 soft tissue, hydroxyapatite, calcifications, water, lipid, carbohydrate, cell nucleus, cholesterol, protein, and urinary stones. 
Electron density $\rho_{\rm e}$ is calculated as
\begin{eqnarray}
\rho_{\rm e} = \rho_{\rm m} \sum_{i} w_i \frac{Z_i}{{A_{\rm r}}_i} \left(\frac{Z}{A_{\rm r}}\right)_{\rm w}^{-1},
\end{eqnarray}
where $Z_i$ and ${A_{\rm r}}_i$ are the atomic number and the atomic weight of element $i$, $(Z/A_{\rm r})_{\rm w} = 0.5551$ is the mean $Z/A_{\rm r}$ of water, and $w_i$ and $\rho_{\rm m}$ are the elemental mass fraction and the mass density of the material.
We represent $\rho_{\rm m}$, $\rho_{\rm e}$, and all other densities as non-dimensional ratios to those of water in this study.

\subsection{Stopping power}

The Bethe theory leads to the stopping-power ratio of a material to water,\cite{ICRU 1993} or the stopping effective density, as
\begin{eqnarray}
\rho_{\rm S} &=& \rho_{\rm e} \left(\ln\frac{2 m_{\rm e} c^2}{I} + \ln\frac{v^2}{c^2-v^2} -\frac{v^2}{c^2} \right)
\nonumber\\*
&\times& 
\left(\ln\frac{2 m_{\rm e} c^2}{I_{\rm w}} + \ln\frac{v^2}{c^2-v^2} -\frac{v^2}{c^2} \right)^{-1},
\label{eq:2}
\end{eqnarray}
where $m_{\rm e} = 0.511$ MeV/$c^2$ is the electron mass, $v$ and $c$ are the speeds of the projectile and light, and $I$ and $I_{\rm w}$ are the mean excitation energies of the material and water.
We calculated the $I$ values of the body tissues with the Bragg rule
\begin{eqnarray}
\ln I = \sum_i w_i \frac{Z_i}{{A_{\rm r}}_i} \ln I_i \left( \sum_i w_i \frac{Z_i}{{A_{\rm r}}_i} \right)^{-1},
\end{eqnarray}
where elemental $I_i/{\rm eV}$ values were H:19.2, C:81, N:82, O:106, F:112, Na:168, Mg:176, P:195, Cl:180, K:215, Ca:216, and Fe:323 for constituents of solid or liquid compounds.\cite{ICRU 1984}
This leads to $I_{\rm w} = 75.3$ eV for water.
Although recent studies suggest slightly higher $I_{\rm w}$ values,\cite{Sigmund \etal 2009} we used this $I_{\rm w}$ value in the denominator of Eq.~(\ref{eq:2}) to cancel out the common systematic error of the compound $I$ value in the numerator for general body tissues whose main ingredient is water.
As the $v$-dependent variation of the $\rho_{\rm S}$ is within 1\% under therapeutic conditions,\cite{Kanematsu \etal 2003} we took the representative projectile speed $v = 0.6\,c$ or the nucleon kinetic energy $E/A = 230$ MeV to define the stopping effective density as projectile independent.

\subsection{Multiple scattering}
Scattering power $T$ is the increase of Gaussian angular variance per transport distance.\cite{ICRU 1984}
In some formulations, it is in the form of
\begin{eqnarray}
T = \left(\frac{E_{\rm s}}{p\, v}\right)^2 \frac{1}{X},
\end{eqnarray}
where $E_{\rm s} = 15.0$ MeV is a constant energy, $p$ is the particle momentum, and $X$ is the scattering length of the material. 
Gottschalk derived the scattering length for heavy charged particles.\cite{, Gottschalk 2010} For a composite material, it is given by
\begin{eqnarray}
\frac{1}{X} = \frac{\rho_{\rm m}}{4298.4\ {\rm cm}} \sum_i w_i \frac{Z_i^2}{{A_{\rm r}}_i} \left( 29.733 - \ln Z_i - \ln {A_{\rm r}}_i \right).
\end{eqnarray}
This leads to $X_{\rm w}= 46.88$ cm for water.
We define the scattering effective density as the scattering-power ratio of the material to water,
\begin{eqnarray}
\rho_{\rm T} = \frac{T}{T_{\rm w}} = \frac{X_{\rm w}}{X}.
\end{eqnarray}

\subsection{Nuclear interactions}

Sihver \etal made an empirical modification to the geometric model for proton--nucleus and nucleus--nucleus collision cross section.\cite{Sihver \etal 1993, Sihver and Mancusi 2009}
Considering the symmetry between target hydrogen and projectile proton, the cross section is written as
\begin{eqnarray}
\sigma_{\rm N} &=& \pi r_0^2 \left[ A^{1/3}+A_{\rm r}^{1/3}-b_0 \left(A^{-1/3}+A_{\rm r}^{-1/3} \right) \right]^2,
\\
b_0 &=& \begin{cases}
2.247 -0.915 \left(A^{-1/3}+A_{\rm r}^{-1/3}\right)\\
\qquad\qquad\qquad\qquad \text{for } A = 1 \text{ or } A_{\rm r} \approx 1,\\
1.581-0.876 \left(A^{-1/3}+A_{\rm r}^{-1/3}\right)\\
\qquad\qquad\qquad\qquad \text{for } A \geq 2 \text{ and } A_{\rm r} \gtrsim 2,
\end{cases} \label{eq:8}
\end{eqnarray}
where $A$ is the mass number of the projectile nucleus, $A_{\rm r}$ is the atomic weight of the target element, $r_0 = 1.36$~fm is the effective nucleon radius, and $b_0$ is the overlap parameter.
In their model, the energy dependence of the cross section would be almost common among target body-tissue materials in the therapeutic energy region and was thus ignored in this study.

For a compound material of mass density $\rho_{\rm m}$, the number of element-$i$ nuclei per volume is proportional to $\rho_{\rm m} w_i /{A_{\rm r}}_i$.
The total nuclear cross section per volume is thus proportional to $\sum_i {\sigma_{\rm N}}_i \rho_{\rm m} w_i /{A_{\rm r}}_i$, which is proportional to the incidence of nuclear interactions. 
We define the nuclear effective density as the ratio of the incidence of nuclear interactions in the material to that in water, resulting in
\begin{eqnarray}
&&\rho_{\rm N} = \sum_i \frac{{\sigma_{\rm N}}_i}{{A_{\rm r}}_i}\, w_i\, \rho_{\rm m} \, \left( \frac{\sigma_{\rm N}}{A_{\rm r}} \right)_{\rm w}^{-1}, \\
&&\left(\frac{\sigma_{\rm N}}{A_{\rm r}}\right)_{\rm w} =\frac{ 2\, {\sigma_{\rm N}}_{\rm H} + {\sigma_{\rm N}}_{\rm O}}{{M_{\rm r}}_{\rm w}},
\end{eqnarray}
where ${\sigma_{\rm N}}_{\rm H}$ and ${\sigma_{\rm N}}_{\rm O}$ are the cross sections of hydrogen and oxygen nuclei and ${M_{\rm r}}_{\rm w} = 18.015$ is the molecular weight of water.
Unlike the other effective densities, we formulated the nuclear effective density as projectile dependent. 

\subsection{Example of corrective conversions}

To demonstrate the procedure and to evaluate the significance of the interaction-specific corrective conversion, we carried out simplistic proton and carbon-ion beam-transport calculations in one-dimensional geometries, each with and without the corrective conversions.
We applied a proton beam of range 9.8 cm in water to a lung-like geometry consisting of staked slabs of 1-cm soft tissue, 2-cm bone, 1-cm soft tissue, 8-cm lung, and 3-cm tumor.
We applied a carbon-ion beam of range 22.4 cm in water to a prostate-like geometry consisting of stacked slabs of 6-cm soft tissue, 6-cm bone, 4-cm soft tissue, and 4-cm tumor.
We assigned electron densities 1.0 for soft tissue and tumor, 1.4 for bone, and 0.25 for lung and set the beam ranges to coincide with the tumor depths without the correction.
These beams were point-like mono-directional at the phantom entrance and were transported through the slabs at 1-mm intervals with quantities of interest,
\begin{eqnarray}
d(x) &=& \int_0^x \rho (x') \, \rmd x'
\\
\sigma_y (x) &=& \left( \int_0^x (x-x')^2 \rho(x')\, T_{\rm w}(x')\, \rmd x' \right)^{1/2}, \label{eq:12}
\end{eqnarray}
evaluated as a function of transport distance $x$, where generic depth, beam size, and density, $(d, \sigma_y, \rho)$, represent the effective depths and the beam sizes calculated with the respective effective densities.
For the integration, the Fermi-Eyges theory in the numerical form\cite{Kanematsu 2008} was implemented with a semi-empirical scattering power for water, $T_{\rm w}$.\cite{Kanematsu 2009}
We are interested in the beam size corrected for multiple scattering and the effective depth corrected for nuclear interactions, $({\sigma_y}_{\rm T}, d_{\rm N})$, to compare to the effective depth and beam size in the conventional heterogeneity correction, $(d_{\rm S}, {\sigma_y}_{\rm S})$.
The effective depths are geometric parameters that will not depend on beam properties except for the $d_{\rm N}$.
We refer to the effective depths of the distal end of the target as the effective target depths $d_{\rm S,tgt}$, $d_{\rm T,tgt}$, and $d_{\rm N,tgt}$ for the individual interactions.

\section{Results}

\subsection{Corrective conversion functions}

\begin{figure}
\includegraphics[width=8.5 cm]{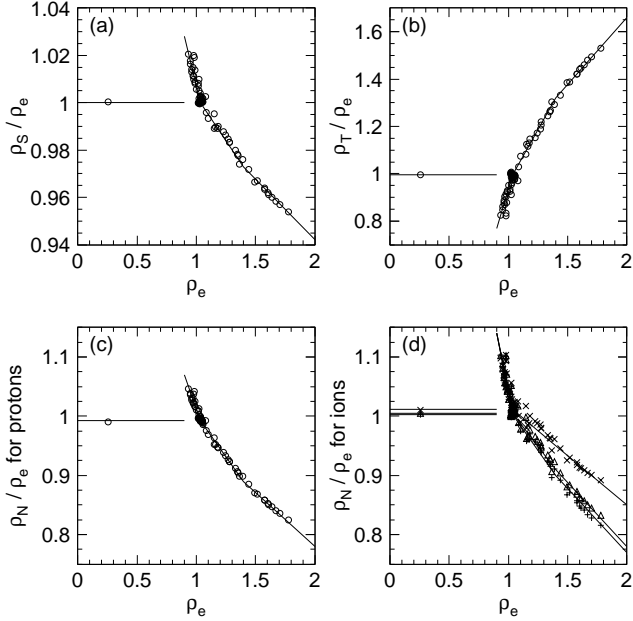}
\caption{Ratios of (a) stopping effective density $\rho_{\rm S}$, (b) scattering effective density $\rho_{\rm T}$, (c) nuclear effective density $\rho_{\rm N}$ for protons, and (d) nuclear effective density $\rho_{\rm N}$ for helium ions ($\times$), carbon ions ($\triangle$), and oxygen ions ($+$) to electron density $\rho_{\rm e}$ with the polyline conversion functions (solid lines).}
\label{fig:1}
\end{figure}

Figure~\ref{fig:1} shows correspondences between the electron density and the effective densities of body tissues for proton and ion beams.
There was a concentration of tissues around $(\rho_{\rm e},\rho_{\rm S}/\rho_{\rm e},\rho_{\rm T}/\rho_{\rm e}, \rho_{\rm N}/\rho_{\rm e}) = (1.035, 1.000, 0.995, 1.00)$.
The low density ($\rho_{\rm e} = 0.258$) for the lung tissue was attributed to the air content.
Generally in the $\rho_{\rm e} \gtrsim 1$ region, a weak negative correlation between $\rho_{\rm S}/\rho_{\rm e}$ and $\rho_{\rm e}$ was observed.
That was attributed to the low $I$ values of carbon-rich adipose tissues in the low $\rho_{\rm e}$ region and the high $I$ values of calcium-rich bone tissues in the high $\rho_{\rm e}$ region.
Simplistically for a single-element material, Coulomb scattering depends on $\rho_{\rm m} Z^2 / A_{\rm r}$ whereas the electron density depends on $\rho_{\rm m} Z / A_{\rm r}$.
As there is approximate proportional relationship among $\rho_{\rm m}$, $A_{\rm r}$, $Z$, and $\rho_{\rm e}$ for common materials, the ratio $\rho_{\rm T}/\rho_{\rm e} \propto Z$ showed a strong positive correlation with $\rho_{\rm e}$.
Similarly, because the nuclear effective density depends approximately on $\rho_{\rm m} {A_{\rm r}}^{-1/3}$ for large $A_{\rm r}$, the ratio $\rho_{\rm N}/\rho_{\rm e} \propto {A_{\rm r}}^{2/3}/Z$ showed a negative correlation with $\rho_{\rm e}$.
Incidentally, while protons, carbon ions, and oxygen ions are similar in the $\rho_{\rm N}/ \rho_{\rm e}$ ratio, helium ions largely deviated from the others.

For the conversion functions, we set a discontinuity point at $\rho_{\rm e} = 0.9$, where none of these tissues existed, an inflection point at the center of the tissue concentration, and another inflection point at $\rho_{\rm e} = 1.4$ in the bone-tissue region.
Similarly to the lung tissues, tissue-air mixing may occur on the surface of skin and epithelial tissues due to finite spatial resolution of the CT image, for which we assigned constant $\rho/\rho_{\rm e}$ ratios in the $\rho_{\rm e} < 0.9$ region.
Although material mixing may generally confuse any conversion schemes, its influence is limited as there normally are only a few tissue boundaries along the beam path.
Table~\ref{tab:1} shows the resultant conversion factors as polyline functions, which are also shown in Fig.~\ref{fig:1}.

The remarkable deviation of helium ions in nuclear interactions could have originated from the limitation of experimental data to construct the Sihver formula.
We thus propose a generic conversion function for the nuclear effective density for protons and ions as a polyline of reduced significant digits, which we used in the examples of lung-like and prostate-like geometries.

\begin{table}
\caption{Polyline conversion functions from electron density $\rho_{\rm e}$ to stopping effective density $\rho_{\rm S}$, scattering effective density $\rho_{\rm T}$, and nuclear effective density $\rho_{\rm N}$.}
\begin{tabular}{lllllll}
\hline\hline
$\rho_{\rm e}$ & 0 & 0.9 & 0.9 & 1.035 & 1.4 & 2.0 \\
\hline
$\rho_{\rm S}/\rho_{\rm e}$ & 1.000 & 1.000 & 1.028 & 1.000 & 0.973 & 0.942 \\
$\rho_{\rm T}/\rho_{\rm e}$ & 0.995 & 0.995 & 0.77 & 0.995 & 1.32 & 1.66 \\
$\rho_{\rm N}/\rho_{\rm e}$ for protons & 0.992 & 0.992 & 1.07 & 0.992 & 0.89 & 0.78 \\
$\rho_{\rm N}/\rho_{\rm e}$ for He ions & 1.011 & 1.011 & 1.14 & 1.011 & 0.95 & 0.85 \\
$\rho_{\rm N}/\rho_{\rm e}$ for C ions & 1.005 & 1.005 & 1.14 & 1.005 & 0.91 & 0.78 \\
$\rho_{\rm N}/\rho_{\rm e}$ for O ions & 1.003 & 1.003 & 1.14 & 1.003 & 0.90 & 0.77 \\
$\rho_{\rm N}/\rho_{\rm e}$ (generic) & 1.0 & 1.0 & 1.1 & 1.0 & 0.9 & 0.8 \\
\hline\hline
\end{tabular}
\label{tab:1}
\end{table}

\subsection{Example of corrective conversions}

\begin{figure}
\includegraphics[width=8.5 cm]{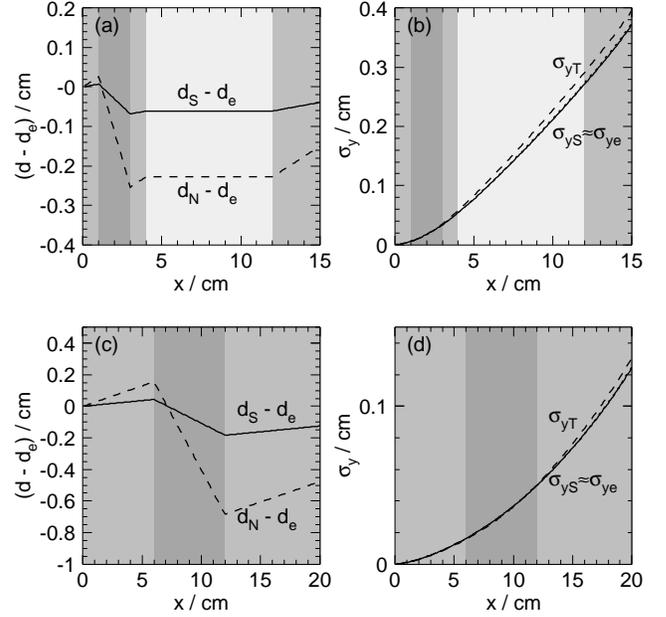} 
\caption{Corrections of stopping and nuclear effective depths, ($d_{\rm S}-d_{\rm e}$, $d_{\rm N}-d_{\rm e}$), and uncorrected, conventional, and corrected beam sizes, (${\sigma_y}_{\rm e}$, ${\sigma_y}_{\rm S}$, ${\sigma_y}_{\rm T}$), with background colors indicating electron densities,  (a, b) for a proton beam in a lung-like geometry and (c, d) for a carbon-ion beam in a prostate-like geometry.}
\label{fig:2}
\end{figure}

Figures \ref{fig:2}(a) and \ref{fig:2}(c) show the development of the depth corrections for stopping and nuclear interactions, $(d_{\rm S}-d_{\rm e})$ and $(d_{\rm N}-d_{\rm e})$, for the proton beam in the lung-like geometry and for the carbon-ion beam in the prostate-like geometry.
The corrections of the stopping effective depth, which would assure the range accuracy, were within about $\pm 0.1$ cm in both cases. 
The bone thickness mainly determined the corrections of the effective target depths. 
The nuclear effective target depth ($d_{\rm N,tgt} = 19.52$ cm) deviated from the stopping effective target detph ($d_{\rm S,tgt} = 19.88$ cm) by $-0.36$ cm in the prostate-like geometry. 

Similarly, figures \ref{fig:2}(b) and \ref{fig:2}(d) show the development of the beam sizes.
The uncorrected beam size ${\sigma_y}_{\rm e}$ and the conventional beam size ${\sigma_y}_{\rm S}$ were almost indistinguishable in these cases, while the corrected beam size ${\sigma_y}_{\rm T}$ deviated from them by about 5\% for the proton beam in the lung-like geometry.
The bone thickness to scatter and the distance to travel determined the amount of the correction. 

\section{Discussion}

A distinctive feature of this study is the generality of the conversion functions.
For example, Szymanowski and Oelfke constructed a conversion function for the angular scaling factor or $\sqrt{\rho_{\rm T}/\rho_{\rm S}}$ in our notation.\cite{Szymanowski and Oelfke 2003} 
While their function was only valid for a specific CT system, the $\rho_{\rm e}$--$\rho_{\rm T}$ function in Table \ref{tab:1} is generally valid.
As the biological data and the physical models are common or equivalent, the direct CT-number conversion and the corrective conversion through $\rho_{\rm e}$ are essentially equivalent in accuracy, assuming that the corrective conversion is used with a stoichiometrically constructed CT-number--$\rho_{\rm e}$ table.\cite{Schneider \etal 1996, Kanematsu \etal 2003}
Intrinsic error of CT number, such as that from x-ray hardening effect (typically 1\% for soft tissues and 2\% for bones),\cite{Schaffner \etal 1998} influences both conversion schemes equally.

Matsufuji \etal found that, for a typical bone of $\rho_{\rm e} \approx 1.35$, the errors of the conventional heterogeneity correction would be $-10\%$ in multiple-scattering angle and $-10\%$ in mean free path of proton--nucleus interaction.\cite{Matsufuji \etal 1998}
For a tissue of $\rho_{\rm e} = 1.35$, the polyline functions in table~\ref{tab:1} give $\rho_{\rm T}/\rho_{\rm e} = 1.21$ and $\rho_{\rm N}/\rho_{\rm e} = 0.904$ for protons.
Because the multiple-scattering angle is proportional to $\sqrt{\rho_{\rm T}}$ and the mean free path is proportional to $1/\rho_{\rm N}$, the $\rho_{\rm T}$ and $\rho_{\rm N}$ conversions would make $+10\%$ and $+11\%$ corrections to multiple-scattering angle and mean free path, respectively.
The $+10\%$ and $+11\%$ corrections correspond to 9 \% and 10\% effects in the corrected values, which agree with their original estimations. 
They also simulated realistic situations, in which bone was not dominant in the beam path, and concluded that the effects on patient dose distributions would be small, which we confirmed in our examples.

The nuclear interactions cause attenuation of primary particles. 
Lee \etal proposed an approximation formula for proton fluence $\Phi(R) = \Phi(0)\left(1+0.012 R/{\rm cm}\right)$ as a linear function of residual range $R$ in water,\cite{Lee \etal 1993} to which the nuclear effective residual range $R_{\rm N} = \int_0^R (\rho_{\rm N}/\rho_{\rm S}) \rmd R'$ should be applied.
For the prostate-like geometry, the error of the nuclear effective target depth with the conventional heterogeneity correction ($\Delta d = d_{\rm S,tgt}-d_{\rm N,tgt} = 0.36$ cm) would cause a relative error of $-0.4\%$ to the proton survival $\Phi(0)/\Phi(R)$ and consequently to the dose to the tumor.
Palmans and Verhaegen reported up to 0.5\% dosimetric effects for their examples.\cite{Palmans and Verhaegen 2005}
For carbon-ion beams, the dosimetric effect would be a few or more times larger for their large nuclear cross section, {\it i.e.} a few percent, which could be clinically influential.

\begin{figure}
\includegraphics[width=8.5 cm]{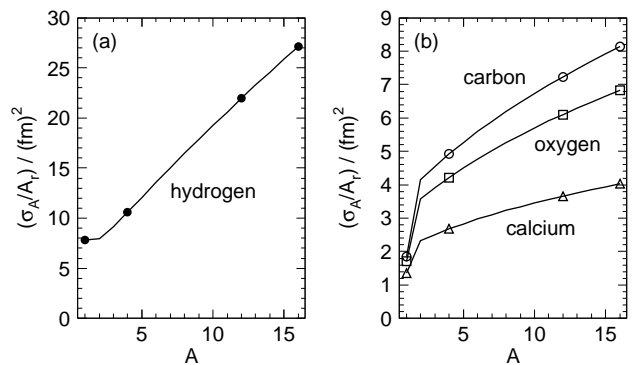}
\caption{Nuclear cross sections per atomic mass unit of hydrogen, carbon, oxygen, and calcium targets as a function of projectile mass number with marker points for those used in this study, calculated with the Sihver formula.}
\label{fig:3}
\end{figure}

Accuracy of the nuclear-interaction model used in this study remains as a concern.
In particular, the distinction of proton (hydrogen ion) from other ions in the the Sihver formula in Eq.~(\ref{eq:8}) seems unnatural and the validity of using the proton--nucleus formula for ion--hydrogen collisions is not very obvious.
Figure~\ref{fig:3} shows the cross sections per atomic mass unit of dominant body-tissue elements for various projectiles.
While the proton ($A=1$) cross sections may be based on abundant data, the cross sections of helium and lighter ions ($A \lesssim 4$) may be probably less reliable, which might have caused the unnatural kinks at $A=2$.

The fragmentation processes in ion beams are further complex.
In the participant--spectator model, the target nucleus will not directly influence how the projectile nucleus may break up.\cite{Baur \etal 1984} 
There are some experimental data for the relative yields of projectile fragments,\cite{Matsufuji \etal 2003} which may be thus reasonably invariant to target materials.
Due to experimental difficulties in precise measurement of a treatment-beam spectrum, Monte Carlo simulation will be useful for building a detailed beam model in numerical or analytical form.\cite{Kempe and Brahme 2010}

\section{Conclusions}

For general body tissues, we found strong correlations between electron density and the effective densities that characterize the strengths of stopping power, multiple scattering, and nuclear interactions of the projectile protons and ions.
The stopping effective density deviated from the electron density by up to a few percent and the scattering and nuclear effective densities deviated by up to a few tens percent, which are consistent with other studies.
To correct those errors, invariant corrective conversion functions from the electron density to the individual effective densities were constructed.
Their effect on patient dose may be negligibly small in general, although it could be at the level of a few percent when a beam traverses a large bone.

The electron density should be derived from the CT number with a conversion table that has been constructed with the best accuracy, {\it i.e.} stoichiometrically, for treatment planning of proton and ion-beam therapy.
The proposed extension of heterogeneity correction will enable improved beam dose calculation without seriously sacrificing simplicity or efficiency of the patient model and the algorithms, while quality-management tasks for CT and treatment panning systems will be kept as they are in the current clinical practice.

\end{document}